\documentclass[cameraready]{Interspeech}

\title{CoSTA: Cognitive-State-Conditioned TTS Data Augmentation Using ASR Transcripts for Alzheimer’s Disease Detection\vspace{-0.6em}}

\author[affiliation={1}]{Yin-Long}{Liu}
\author[affiliation={2}]{Yuanchao}{Li}
\author[affiliation={1}]{Yiming}{Wang}
\author[affiliation={1}]{Yue}{Li}
\author[affiliation={1}]{Rui}{Feng}
\author[affiliation={1}]{Jiaxin}{Chen}
\author[affiliation={1}]{Shaobo}{Liu}
\author[affiliation={1}]{Liu}{He}
\author[affiliation={1}]{Yuang}{Chen}
\author[affiliation={1}, correspondingauthor]{Jiahong}{Yuan}
\author[affiliation={1}]{Zhen-Hua}{Ling}

\address{
 \vspace{-0.7em}
    $^1$ NERC-SLIP, University of Science and Technology of China, Hefei, China \\
    $^2$ Centre for Speech Technology Research, University of Edinburgh, Edinburgh, UK 
}

\email{lyl2001@mail.ustc.edu.cn, \{jiahongyuan, zhling\}@ustc.edu.cn}


\usepackage{cite}
\usepackage{hyperref}
\hypersetup{
    colorlinks=false,
    pdfborder={0 0 1}
}
\usepackage{amsmath,amssymb,amsfonts}

\usepackage{algorithmic}
\usepackage{graphicx}
\usepackage{textcomp}
\usepackage{marvosym}        
\usepackage{xcolor}
\def\BibTeX{{\rm B\kern-.05em{\sc i\kern-.025em b}\kern-.08em
    T\kern-.1667em\lower.7ex\hbox{E}\kern-.125emX}}
\usepackage{amsmath,graphicx,hyperref}
\usepackage{marvosym}        
\usepackage{booktabs,multirow}
\usepackage{array}
\usepackage{siunitx}
\usepackage{tikz}
\usetikzlibrary{arrows.meta,positioning,fit,shapes}
\usepackage{threeparttable}
\usepackage{tabularx}
\usepackage{cite}
\usepackage{enumitem}
\usepackage[ruled,vlined]{algorithm2e} 
\usepackage{CJKutf8} 
\usepackage{pifont}
\usepackage[svgnames]{xcolor} 
\usepackage{amsmath}
\usepackage{diagbox} 
\usepackage{colortbl} 
\definecolor{ForestGreen}{rgb}{0.13, 0.55, 0.13}
\definecolor{DeepForestGreen}{rgb}{0.13, 0.45, 0.13}
\definecolor{FireBrick}{rgb}{0.7, 0.13, 0.13}
\definecolor{Gray}{gray}{0.95}
\newcommand{\up}[1]{\rlap{\textcolor{FireBrick}{\fontsize{6pt}{0pt}\selectfont$_{\uparrow#1}$}}}
\newcommand{\down}[1]{\rlap{\textcolor{ForestGreen}{\fontsize{6pt}{0pt}\selectfont$_{\downarrow#1}$}}}

\begin{document}
\maketitle
\begin{abstract}
Speech-based Alzheimer's Disease (AD) detection is constrained by scarce pathological speech data. To address this, we propose CoSTA, a Text-to-Speech (TTS)-based data augmentation framework. Specifically, we first develop two Cognitive-State-Conditioned (CS-Cond) TTS models by adapting CosyVoice2 and F5-TTS to synthesize speech with distinct AD and Healthy Control characteristics. Furthermore, by constructing a transcript pool comprising Manual Transcripts (MT) and 36 Automatic Speech Recognition (ASR) transcripts, we investigate the impact of text sources on TTS-based augmentation. We also perform augmentation-factor analysis and test-time augmentation. Experiments on the ADReSS dataset show that CS-Cond TTS significantly improves synthetic speech utility, and ASR-driven augmentation frequently outperforms MT-driven augmentation. Finally, CoSTA yields a 4.16\% gain over the baseline, achieving an audio-only accuracy of 85.83\% on the ADReSS test set and outperforming prior methods.
\end{abstract}

\begin{figure*}[t!]
  \centering
  \includegraphics[width=0.9\linewidth]{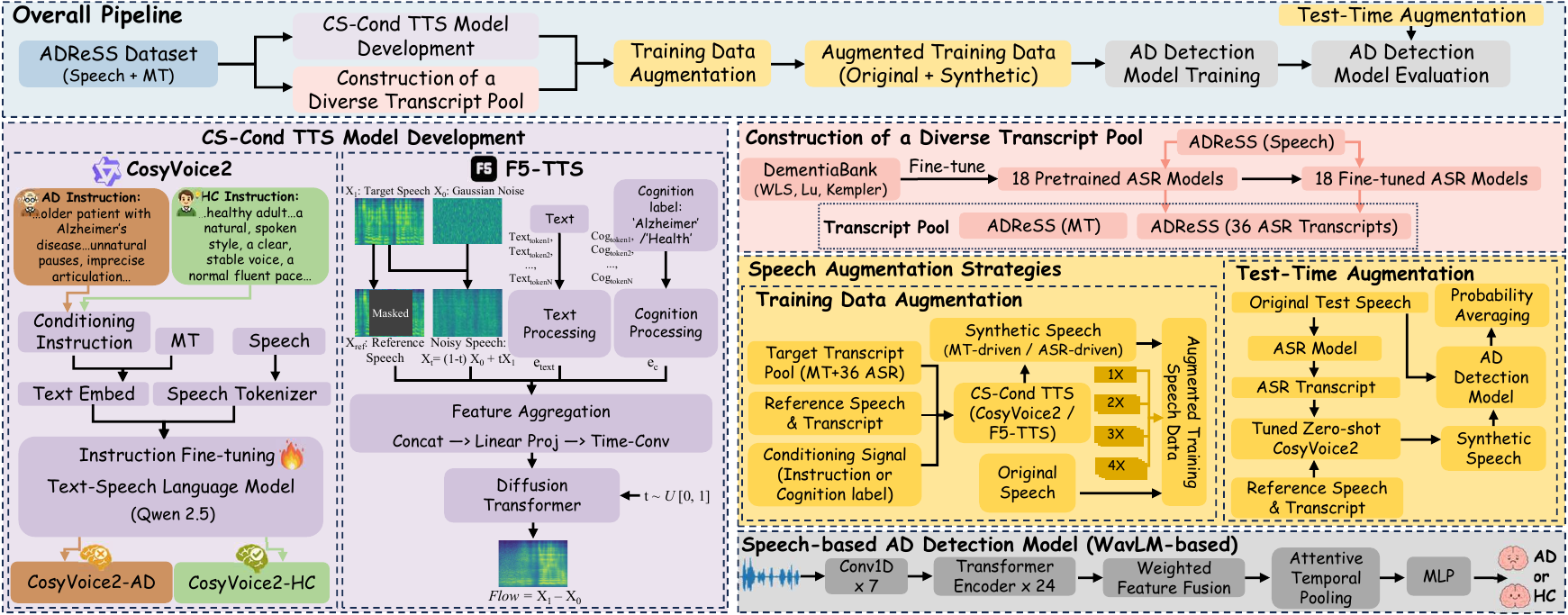} 
  \caption{Overview of the proposed CoSTA framework.}
  \label{fig:pipeline}
  \vspace{-0.6cm}
\end{figure*}

\vspace{-0.2cm}
\section{Introduction}
\vspace{-0.1cm}
\label{sec:Introduction} 
Alzheimer's Disease (AD) causes irreversible cognitive decline, impairing memory, language, and executive function. While early detection is critical, conventional diagnostics like Positron Emission Tomography are costly and invasive \cite{de2019brief}. Since speech production couples cognition and motor control, AD manifests as temporal disfluencies, imprecise articulation, and lexical deficits \cite{yang2022deep}. Consequently, speech-based AD detection is a promising screening tool to distinguish AD from Healthy Controls (HC) \cite{li2025semi,liu2025can,li2025detecting,liu2026cross,mei2023ustc}. However, developing robust AD detection models is severely constrained by data scarcity from privacy regulations and limited patient availability. This sparsity makes models prone to overfitting and poor generalization, highlighting the critical need for effective Data Augmentation (DA).

While prior studies have explored speech DA for AD detection \cite{hason2022spontaneous, woszczyk22_interspeech}, the majority rely on traditional signal-level perturbations (e.g., noise injection, pitch shifting, and time stretching). 
These approaches primarily generate distorted variants of existing recordings, without introducing new semantic content or explicitly modeling pathology-specific speaking traits, such as AD-like unnatural pauses, which serve as critical discriminative cues for AD detection.
Moreover, such indiscriminate perturbations may even degrade detection performance \cite{woszczyk22_interspeech}.

Recent advances in Text-to-Speech (TTS) offer a promising alternative for DA. By converting text into speech, TTS can generate vast training samples with diverse acoustic traits, a strategy proven effective in data-scarce Automatic Speech Recognition (ASR) settings \cite{leung2025text}. However, TTS-based DA remains largely unexplored for AD detection. A critical challenge is the mismatch between standard TTS objectives and AD diagnostic requirements: standard TTS models are typically optimized for intelligibility and naturalness~\cite{du2024cosyvoice, chen2025f5}, which inherently regularizes disfluencies and prosodic irregularities, thereby masking acoustic biomarkers of AD. Consequently, \textbf{we argue that effective DA must leverage Cognitive-State-Conditioned (CS-Cond) TTS to intentionally synthesize speech reflecting distinct AD-like versus HC-like characteristics.}

Another underexplored yet critical dimension is the choice of text sources used to drive TTS. While Manual Transcripts (MT) provide the ground truth, recent studies suggest that ASR errors may encode diagnostic cues that occasionally benefit text-level AD detection over perfect transcripts \cite{LI2024104598}. This raises a fundamental question for TTS-based DA: \textit{Should synthetic speech be driven by MT or ASR transcripts?} Unlike MT, ASR transcripts introduce diverse error patterns and linguistic perturbations, increasing data variance and potentially providing synthetic speech with diagnostically relevant pathological cues.

To address these challenges, we propose CoSTA, a \textbf{Co}gnitive-\textbf{S}tate-Conditioned \textbf{T}TS-based data \textbf{A}ugmentation framework. Our main contributions are as follows:
\begin{itemize}[noitemsep, topsep=0pt, leftmargin=*]
    \item We develop two CS-Cond TTS models by adapting CosyVoice2 and F5-TTS, enabling controlled synthesis of AD-like versus HC-like speech for effective DA.
    \item We systematically investigate the impact of text sources on TTS-based DA by constructing a transcript pool comprising MT and 36 ASR transcripts, revealing that ASR-driven augmentation frequently outperforms its MT-driven counterpart.
    \item Extensive experiments on the ADReSS dataset demonstrate CoSTA's efficacy. It achieves an audio-only accuracy of 85.83\% on the test set, surpassing prior methods.
\end{itemize}

\vspace{-0.2cm}
\section{Methods}
\vspace{-0.1cm}
The CoSTA framework is illustrated in Figure~\ref{fig:pipeline}. This section outlines its four components: CS-Cond TTS model development, construction of a diverse transcript pool, speech augmentation strategies, and the speech-based AD detection model.
\vspace{-0.1cm}
\subsection{CS-Cond TTS model development}
\vspace{-0.1cm}
\subsubsection{CS-Cond CosyVoice2}
\vspace{-0.1cm}
CosyVoice2~\cite{du2024cosyvoice} employs a pretrained LLM (Qwen2.5)~\cite{team2024qwen2} as a unified text--speech Language Model (LM) to autoregressively generate discrete speech tokens. To achieve cognitive controllability, we leverage its instruction fine-tuning capability. As shown in Figure~\ref{fig:pipeline}, we design natural-language instructions $\mathcal{I}^{(c)}$ for each cognitive state $c\in\{\mathrm{AD},\mathrm{HC}\}$. During fine-tuning, an instruction is concatenated with the target MT $\mathbf{y}_{\mathrm{MT}}$ using a separator \texttt{<|endofprompt|>} to form a unified text prompt:  
\vspace{-3mm} 
\begin{equation}
\mathbf{y}^{(c)} = [\mathcal{I}^{(c)},\ \texttt{<|endofprompt|>},\ \mathbf{y}_{\mathrm{MT}}]
\end{equation}
\vspace{-6mm} 

The text embed module converts $\mathbf{y}^{(c)}$ into tokens $\mathbf{t}^{(c)}=\{t_1,\dots,t_M\}$, while paired training speech is discretized by the speech tokenizer into $\mathbf{s}=\{s_1,\dots,s_T\}$. 
We fine-tune the text--speech LM by minimizing the negative log-likelihood of ground-truth speech tokens $\mathbf{s}$ conditioned on the text prompt:
\vspace{-2.5mm} 
\begin{equation}
\label{eq:cosyvoice_loss}
\mathcal{L}_{\mathrm{LM}} = - \sum_{k=1}^{T} \log P_{\theta}\big( s_k \bigm| \mathbf{t}^{(c)}, \mathbf{s}_{<k} \big)
\end{equation}
\vspace{-3.5mm}

where $\theta$ denotes the LM parameters. By performing this fine-tuning on AD-specific and HC-specific subsets of the training data respectively, we obtain two specialized variants: CosyVoice2-AD and CosyVoice2-HC. During inference, a target transcript, instruction $\mathcal{I}^{(c)}$, and class-matched reference speech are fed into the respective model to generate speech tokens. These tokens are converted into Mel-spectrograms via a flow-matching-based reconstruction module \cite{lipman2023flow}, and finally into waveforms using a pretrained HiFi-GAN vocoder \cite{kong2020hifi}.

\vspace{-0.1cm}
\subsubsection{CS-Cond F5-TTS}
\vspace{-0.1cm}
To complement the autoregressive approach, we adapt F5-TTS \cite{chen2025f5}, a non-autoregressive model based on Flow Matching (FM) with a Diffusion Transformer (DiT) backbone \cite{peebles2023scalable}. It models speech generation as transforming Gaussian noise $x_0$ into the target distribution $x_1$ by estimating a velocity field $v_t$.

As illustrated in Figure~\ref{fig:pipeline}, to enable cognitive controllability, we incorporate a Cognition Processing block analogous to the Text Processing block. This block comprises a sequence of ConvNeXtv2 \cite{woo2023convnext} layers with RoPE encoding \cite{su2024roformer} to map a discrete cognition label $l_c \in \{\text{Alzheimer}, \text{Health}\}$ into a dense embedding $\mathbf{e}_c$. A Feature Aggregation module then processes four inputs: the cognition embedding $\mathbf{e}_c$, text embedding $\mathbf{e}_{\text{text}}$, reference Mel-spectrogram $\mathbf{x}_{\text{ref}}$, and noisy Mel-spectrogram $\mathbf{x}_t$. Subsequently, a DiT backbone with RoPE predicts the flow. The model is trained by minimizing the FM loss:
\vspace{-1.5mm} 
\begin{equation}
\mathcal{L}_{\text{FM}} = \mathbb{E}_{t, x_0, x_1} \left\lVert v_{\theta}(t, x_t, \mathbf{C}) - (x_1 - x_0) \right\rVert^2 
\end{equation}
\vspace{-5mm} 

where $\theta$ denotes model parameters, $t \sim \mathcal{U}[0,1]$ is the time step, $x_t = (1-t)x_0 + t x_1$ represents the interpolated noisy state, and $\mathbf{C}$ is the conditioning context. Unlike CosyVoice2, we train a unified CS-Cond F5-TTS on a mixture of AD and HC samples. During inference, a target transcript, label $l_c$, and class-matched reference speech/text are fed into CS-Cond F5-TTS to generate log Mel-spectrograms, which are converted into audio signals via a pretrained Vocos vocoder \cite{siuzdak2024vocos}.

\vspace{-0.1cm}
\subsection{Construction of a diverse transcript pool}
\vspace{-0.1cm}
\label{subsec:transcript_pool}
To systematically investigate the impact of text sources on TTS-based DA, we construct a diverse transcript pool comprising MT and 36 ASR transcripts. Specifically, we first fine-tune 18 pretrained ASR models across four widely used families: Wav2Vec2 \cite{baevski2020wav2vec}, HuBERT \cite{hsu2021hubert}, WavLM \cite{chen2022wavlm}, and Whisper \cite{radford2023robust}. Selected for their diverse architectures and training paradigms to yield varied transcription fidelities and error distributions, these models (available on HuggingFace) include: 
\textit{\textbf{wav2vec2}-\{base-100h, base-960h, large-960h, large-960h-lv60, large-960h-lv60-self, large-xlsr-53-english, xls-r-1b-english\}}, 
\textit{\textbf{hubert}-\{large-ls960-ft, xlarge-ls960-ft\}}, 
\textit{\textbf{wavlm}-libri-clean-100h-\{base-plus, large\}}, 
and \textit{\textbf{whisper}-\{tiny, base, small, medium, large, large-v2, large-v3\}}. After fine-tuning, we utilize the full ensemble of 36 models (18 pretrained and 18 fine-tuned) to transcribe the AD dataset, generating 36 distinct ASR transcripts per speech sample. These transcripts, alongside the MT, are fed into the TTS models as target text, augmenting the original training dataset and enabling comparative analysis between ASR-driven and MT-driven augmentation strategies.

\vspace{-0.1cm}
\subsection{Speech augmentation strategies}
\label{subsec:augmentation_strategies}
\vspace{-0.1cm}
\subsubsection{Training data augmentation}
\vspace{-0.1cm}
We leverage four TTS models (pretrained and CS-Cond variants of CosyVoice2 and F5-TTS) and the transcript pool to augment training data. 
Let the original training dataset be $\mathcal{D} = \mathcal{D}_{\mathrm{AD}} \cup \mathcal{D}_{\mathrm{HC}}$, where $\mathcal{D}_{\mathrm{AD}} = \{(s^{\mathrm{AD}}_i, t^{\mathrm{AD}}_i)\}_{i=1}^{M}$ and $\mathcal{D}_{\mathrm{HC}} = \{(s^{\mathrm{HC}}_j, t^{\mathrm{HC}}_j)\}_{j=1}^{N}$ contain speech $s$ and MT $t$ for $M$ AD and $N$ HC subjects.
For each speech sample $s_k \in \mathcal{D}$,  we have access to a set of 37 transcripts $\mathcal{T}_k = \{t_k\} \cup \{t^{(a)}_k\}_{a=1}^{36}$ (one MT, 36 ASR transcripts). To generate a synthetic sample $\hat{s}_k$ from a target transcript $t_{\mathrm{tar}} \in \mathcal{T}_k$, we employ the TTS model $\Phi$ conditioned on the signal $C_k$, which is consistent with the ground-truth class of $s_k$. The process is formulated as:

\vspace{-4mm}
\begin{equation}
    \hat{s}_k = \Phi(t_{\mathrm{tar}}, C_k, s_{\mathrm{ref}}, t_{\mathrm{ref}})
\end{equation}
\vspace{-5.5mm}

where $s_{\mathrm{ref}}$ and $t_{\mathrm{ref}}$ denote the reference speech and text, respectively. To achieve variable augmentation factors (e.g., $2\times$, $3\times$, $4\times$), we use two strategies:

\begin{itemize}[noitemsep, topsep=0pt, leftmargin=*]
    \item Self-Reference Synthesis ($2\times$): We set $s_{\mathrm{ref}} = s_k$ and $t_{\mathrm{ref}} = t_k$. The synthetic variant $\hat{s}_k$ retains the original speaker's timbre while incorporating the linguistic characteristics of $t_{\mathrm{tar}}$. 
    \item Intra-Class Cross-Synthesis ($>2\times$): To further diversify the training distribution, we randomly sample a reference speech $s_n$ from a different subject $n$ within the same class (i.e., class($k$) = class($n$), $k \neq n$). This combines the linguistic content of subject $k$ with the timbre of subject $n$, allowing us to construct datasets beyond $2\times$ via iterative cross-synthesis.
\end{itemize}
 
\vspace{-0.1cm}
\subsubsection{Test-time augmentation (TTA)}
\vspace{-0.1cm}
To further enhance performance, we implement TTA. Since the ground-truth cognitive classes are unavailable during testing, the CS-Cond generation cannot be explicitly applied. Instead, we fine-tune a zero-shot CosyVoice2 model on the training set without cognitive instruction conditioning. For a given test speech $s_{\mathrm{test}}$, we first transcribe it using the same ASR model employed during the training augmentation to obtain the transcript $t_{\mathrm{asr}}$. We then synthesize a speech variant $\hat{s}_{\mathrm{test}}$ via the fine-tuned zero-shot model, utilizing $s_{\mathrm{test}}$ as the reference speech and $t_{\mathrm{asr}}$ as the target text.
Both the original $s_{\mathrm{test}}$ and the synthetic $\hat{s}_{\mathrm{test}}$ are fed into the trained AD detection model to obtain class probability distributions $\mathbf{P}_{\mathrm{ori}} = [p^{\mathrm{ori}}_{\mathrm{HC}}, p^{\mathrm{ori}}_{\mathrm{AD}}]$ and $\mathbf{P}_{\mathrm{syn}} = [p^{\mathrm{syn}}_{\mathrm{HC}}, p^{\mathrm{syn}}_{\mathrm{AD}}]$, respectively. The final prediction is derived via probability averaging: $\mathbf{P}_{\mathrm{final}} = (\mathbf{P}_{\mathrm{ori}} + \mathbf{P}_{\mathrm{syn}}) / 2$.

\vspace{-0.1cm}
\subsection{Speech-based AD detection model}
\vspace{-0.1cm}
We propose an audio-only AD detection model built upon a pretrained WavLM. Given a 16 kHz speech waveform, a 7-layer strided convolutional feature extractor generates frame-level representations with a downsampling factor of 320. These features are normalized, linearly projected to a 1024-dimensional space, and processed by a 24-layer Transformer encoder. To exploit hierarchical representations, a weighted fusion module combines hidden states from all layers via learnable softmax weights. An attentive temporal pooling module then aggregates these features into a fixed-dimensional vector. Finally, a 3-layer MLP with softmax predicts the binary probabilities (AD vs.\ HC). The model is trained end-to-end using cross-entropy loss.

\begin{figure}[b!]
 \vspace{-0.6cm}
  \centering
  \includegraphics[width=0.7\linewidth]{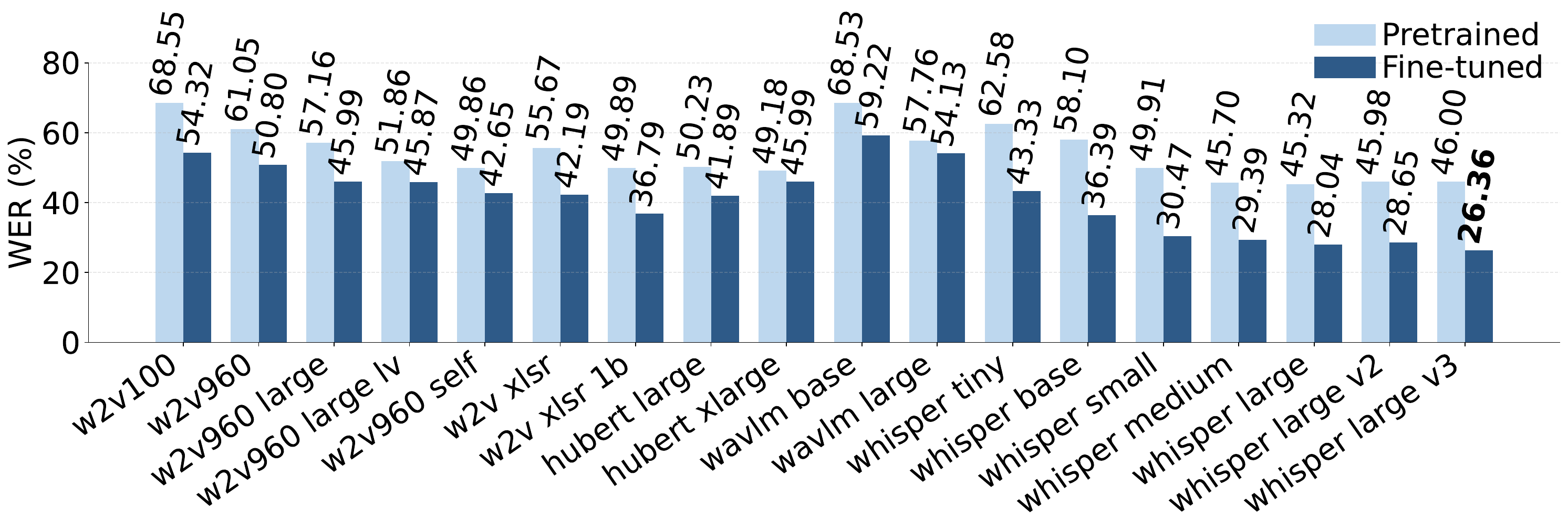} 
  \caption{Mean WER (\%) of 36 ASR models on ADReSS dataset.}
  \label{fig:asr_wer}
\end{figure}

 \vspace{-0.2cm}
\section{Experiments}
\label{sec:Experiments}
 \vspace{-0.1cm}
\subsection{Dataset}
 \vspace{-0.1cm}
For AD detection, we used the ADReSS dataset \cite{luz2020fuente}, which contains a training set of 108 subjects (54 AD, 54 HC; $\approx$\SI{1.7}{h}) and a test set of 48 subjects (24 AD, 24 HC; $\approx$\SI{0.9}{h}). Each subject performed the ``Cookie Theft'' picture description task \cite{goodglass1983boston}, producing a single speech-MT pair. We randomly partitioned the ADReSS training set into a TTS training set (45 AD, 45 HC) and a TTS test set (9 AD, 9 HC) at a 5:1 ratio. Based on the provided timestamp information, we segmented original samples into $\approx$\SI{30}{\second} speech-text pairs for TTS training. To fine-tune the ASR models, we used three DementiaBank subsets (WLS \cite{herd2014cohort}, Lu \cite{lanzi2023dementiabank}, and Kempler \cite{kempler1987syntactic}) comprising 245 samples ($\approx$\SI{3}{h}) from the same picture description task.

\vspace{-0.1cm}
\subsection{Implementation details}
\vspace{-0.1cm}
For training CosyVoice2, we used the Adam optimizer with a lr of $1\times 10^{-5}$ and adopted a dynamic batch strategy, with a maximum of 2,000 frames per batch. We empirically found that training for one epoch was sufficient.
For training F5-TTS, we used the AdamW optimizer with a lr of $1\times 10^{-5}$ and a batch size of 8, training for 40 epochs.
We employed three metrics to objectively evaluate the TTS models: Mel Cepstral Distortion (MCD)~\cite{kominek2008synthesizer}, Log $F_0$ Root Mean Square Error (Log-$F_0$ RMSE)~\cite{wang08_iscslp}, and Frechet Audio Distance (FAD)~\cite{kilgour19_interspeech}. 
Specifically, we synthesized speech using the MT from the TTS test set and paired it with ground-truth speech to calculate these metrics. WavLM embeddings were utilized for FAD calculation.
For fine-tuning ASR models, we used the AdamW optimizer with a lr of $1\times 10^{-5}$ and a batch size of 8, training for 20 epochs, utilizing Word Error Rate (WER) for evaluation. For training the AD detection model, we utilized AdamW optimizer with a lr of $5\times 10^{-5}$ and a batch size of 8, training for 30 epochs, reporting the average accuracy obtained from five independent runs for evaluation. All experiments were conducted on NVIDIA A800 GPUs with 80GB of VRAM. 

\begin{table}[t!]
\centering
\caption{Objective evaluation results on the TTS test set. 
}
\label{tab:tts_objective_metrics}

\renewcommand{\arraystretch}{0.5} 
\setlength{\tabcolsep}{3pt}

\resizebox{0.6\linewidth}{!}{
\begin{tabular}{@{}ccccc@{}}
\toprule
\textbf{TTS model} & \textbf{Variant} & \textbf{MCD} $\downarrow$ & \textbf{log-F0 RMSE} $\downarrow$ & \textbf{FAD} $\downarrow$ \\
\midrule
\multirow{4}{*}{CosyVoice2} & Pretrained-AD & 6.854 & 0.328 & 8.542 \\
                    & CS-Cond-AD  & \textbf{5.436} & \textbf{0.305} & \textbf{2.192} \\
\cmidrule(lr){2-5}
                    & Pretrained-HC & 7.537 & 0.328 & 6.206 \\
                    & CS-Cond-HC  & \textbf{6.933} & \textbf{0.301}  & \textbf{2.459} \\
\midrule
\multirow{4}{*}{F5-TTS} & Pretrained-AD & 6.187 & 0.320 & 2.964 \\
                    & CS-Cond-AD  & \textbf{5.734} & \textbf{0.310} & \textbf{2.545} \\
\cmidrule(lr){2-5}
                    & Pretrained-HC & 7.340 & 0.316  & 3.165 \\
                    & CS-Cond-HC  & \textbf{7.271} & \textbf{0.312} & \textbf{2.744} \\
\bottomrule
\end{tabular}
}
\vspace{-0.7cm}
\end{table}

\vspace{-0.2cm}
\section{Results and analysis}
\vspace{-0.1cm}
\subsection{Objective evaluation of TTS models}
 \vspace{-0.1cm}
Table~\ref{tab:tts_objective_metrics} presents the objective evaluation results of the TTS models. From this table, we observe that: 
\textit{\textbf{1)}} Across all metrics for both AD and HC classes, the CS-Cond models consistently outperform their pretrained counterparts. This confirms that explicitly modeling the cognitive state brings the acoustic profile significantly closer to the ground truth. \textit{\textbf{2)}} CS-Cond CosyVoice2 outperforms CS-Cond F5-TTS across all metrics for both classes, suggesting superior distribution matching.


\begin{figure}[b]
 \vspace{-0.7cm}
  \centering
  \includegraphics[width=0.5\linewidth]{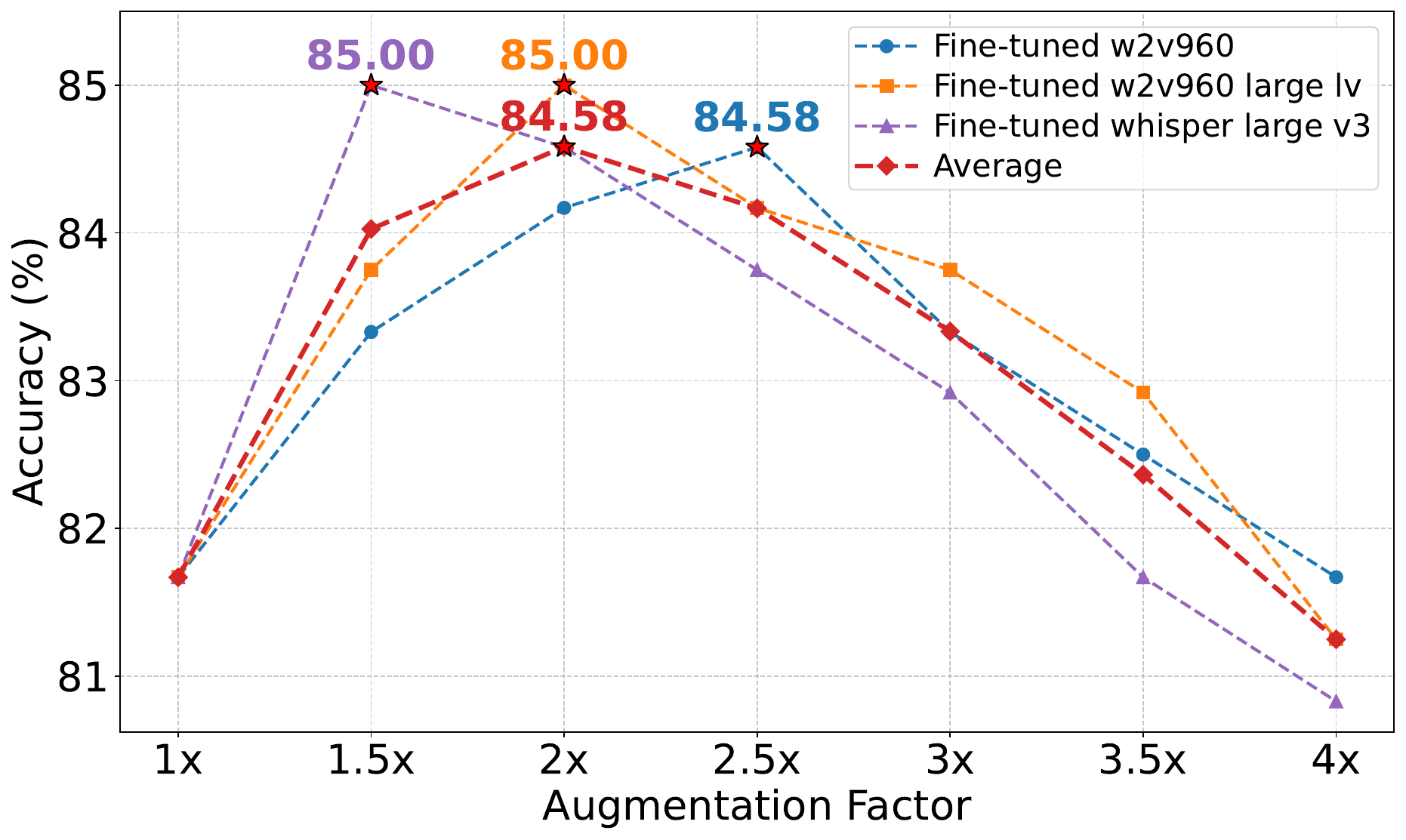} 
    \caption{Impact of augmentation factor on AD detection accuracy. Experiments utilize CS-Cond CosyVoice2 to synthesize speech based on the three high-performing ASR transcripts (marked in \textcolor{SteelBlue}{\textbf{blue}} in Table~\ref{tab:tts_aug_2x_results}).}
  \label{fig:aug_factor}
\end{figure}

 \vspace{-0.1cm}
\subsection{Diversity of ASR transcripts}
 \vspace{-0.1cm}
\label{subsec:ASR_Diversity}
Figure~\ref{fig:asr_wer} illustrates the WER of the 18 ASR models, comparing their pretrained and fine-tuned versions on the ADReSS dataset. As observed, fine-tuning consistently improves ASR performance, yielding lower WER across all models. The WER values span a broad range, ranging from 26.36\% to 68.55\%. This substantial variance in transcription quality provides a diverse pool of ASR transcripts for the TTS-based DA.

\begin{table*}[t]
\centering
\caption{Comparison of AD detection accuracy (\%). The baseline is trained on the original training set (108 samples), while comparison models use a $2\times$ augmented training set (original + synthetic, 216 samples). Synthetic speech is generated by four TTS models using either Manual Transcripts (MT) or one of the 36 ASR transcripts (18 pretrained + 18 fine-tuned). \textcolor{FireBrick}{$\uparrow$} indicates accuracy surpassing the baseline (81.67\%). \textbf{Bold} denotes ASR-driven augmentation outperforming its MT-driven counterpart per TTS model. \textcolor{SteelBlue}{\textbf{Blue bold}} values denote configurations ($>$84\%) selected for further analysis on augmentation factors and test-time augmentation.}

\label{tab:tts_aug_2x_results}
\renewcommand{\arraystretch}{0.35} 
\setlength{\tabcolsep}{3pt} 

\resizebox{0.9\linewidth}{!}{
\begin{tabular}{@{}ccccccccc@{}}
\toprule
\textbf{Baseline (Original training set)} & \multicolumn{8}{c}{81.67} \\
\midrule
\multicolumn{9}{c}{\textit{\textbf{--- 2$\times$ TTS-based Data Augmentation ---}}} \\
\midrule
\diagbox[width=10em]{\textbf{Text source}}{\textbf{TTS model}} & \multicolumn{2}{c}{\textbf{Pretrained CosyVoice2}} & \multicolumn{2}{c}{\textbf{CS-Cond CosyVoice2}} & \multicolumn{2}{c}{\textbf{Pretrained F5-TTS}} & \multicolumn{2}{c}{\textbf{CS-Cond F5-TTS}} \\
\midrule
\midrule
\textbf{Manual transcripts (MT)} & \multicolumn{2}{c}{80.83} & \multicolumn{2}{c}{82.50\up{0.83}} & \multicolumn{2}{c}{81.25} & \multicolumn{2}{c}{82.08\up{0.41}} \\
\midrule
\midrule
\textbf{ASR transcripts} & \textbf{Pretrained} & \textbf{Fine-tuned} & \textbf{Pretrained} & \textbf{Fine-tuned} & \textbf{Pretrained} & \textbf{Fine-tuned} & \textbf{Pretrained} & \textbf{Fine-tuned} \\
\midrule
w2v100 & \textbf{81.67} & \textbf{81.67} & 82.08\up{0.41} & 82.50\up{0.83} & 80.42 & 80.83 & 81.25 & 81.67 \\
w2v960 & \textbf{81.67} & 80.83 & 81.67 & \textcolor{SteelBlue}{\textbf{84.17}}\up{2.50} & 79.58 & \textbf{81.67} & 81.67 & \textbf{82.92}\up{1.25} \\
w2v960 large & \textbf{82.50}\up{0.83} & \textbf{82.08}\up{0.41} & 82.08\up{0.41} & \textbf{82.92}\up{1.25} & \textbf{81.67} & \textbf{81.67} & 81.67 & \textbf{83.33}\up{1.66} \\
w2v960 large lv & \textbf{81.25} & 80.00 & 82.08\up{0.41} & \textcolor{SteelBlue}{\textbf{85.00}}\up{3.33} & \textbf{82.92}\up{1.25} & \textbf{82.92}\up{1.25} & \textbf{83.33}\up{1.66} & \textbf{83.75}\up{2.08} \\
w2v960 self & \textbf{81.25} & \textbf{81.25} & 81.67 & \textbf{82.92}\up{1.25} & 80.83 & \textbf{82.50}\up{0.83} & 81.25 & \textbf{82.92}\up{1.25} \\
w2v xlsr & 80.00 & 80.42 & 81.25 & \textbf{83.33}\up{1.66} & 80.42 & 81.25 & 80.42 & \textbf{82.50}\up{0.83} \\
w2v xlsr 1b & 80.83 & \textbf{81.25} & \textbf{82.92}\up{1.25} & 82.08\up{0.41} & \textbf{82.92}\up{1.25} & \textbf{81.67} & \textbf{82.50}\up{0.83} & 81.67 \\
hubert large & \textbf{81.67} & 80.00 & \textbf{83.33}\up{1.66} & \textbf{82.92}\up{1.25} & 81.25 & \textbf{81.67} & 81.25 & \textbf{82.50}\up{0.83} \\
hubert xlarge & \textbf{82.50}\up{0.83} & 80.42 & \textbf{83.75}\up{2.08} & \textbf{82.92}\up{1.25} & \textbf{81.67} & \textbf{82.50}\up{0.83} & \textbf{82.50}\up{0.83} & \textbf{83.33}\up{1.66} \\
wavlm base & 80.42 & \textbf{82.08}\up{0.41} & 81.25 & \textbf{82.92}\up{1.25} & 80.83 & \textbf{82.08}\up{0.41} & 82.08\up{0.41} & \textbf{83.33}\up{1.66} \\
wavlm large & \textbf{81.25} & \textbf{82.08}\up{0.41} & \textbf{83.33}\up{1.66} & \textbf{83.33}\up{1.66} & \textbf{82.08}\up{0.41} & \textbf{82.08}\up{0.41} & \textbf{83.33}\up{1.66} & \textbf{82.50}\up{0.83} \\
whisper tiny & 78.33 & 80.42 & 81.67 & 81.67 & 80.83 & \textbf{82.08}\up{0.41} & \textbf{82.92}\up{1.25} & 81.67 \\
whisper base & 80.00 & \textbf{81.67} & 80.00 & \textbf{82.92}\up{1.25} & 81.25 & \textbf{82.50}\up{0.83} & 81.67 & \textbf{82.92}\up{1.25} \\
whisper small & 80.42 & 80.00 & 81.67 & 80.42 & \textbf{82.08}\up{0.41} & \textbf{82.08}\up{0.41} & 81.67 & \textbf{82.50}\up{0.83} \\
whisper medium & 80.42 & 80.83 & 82.08\up{0.41} & \textbf{83.33}\up{1.66} & \textbf{82.08}\up{0.41} & 81.25 & \textbf{82.50}\up{0.83} & 81.67 \\
whisper large & 80.42 & \textbf{83.33}\up{1.66} & 82.08\up{0.41} & \textbf{83.33}\up{1.66} & 80.83 & \textbf{82.08}\up{0.41} & 80.83 & \textbf{82.92}\up{1.25} \\
whisper large v2 & \textbf{81.25} & \textbf{82.08}\up{0.41} & \textbf{82.92}\up{1.25} & \textbf{82.92}\up{1.25} & \textbf{82.92}\up{1.25} & \textbf{82.92}\up{1.25} & \textbf{82.50}\up{0.83} & \textbf{83.33}\up{1.66} \\
whisper large v3 & 80.83 & \textbf{81.25} & \textbf{82.92}\up{1.25} & \textcolor{SteelBlue}{\textbf{84.58}}\up{2.91} & 81.25 & \textbf{81.67} & \textbf{82.50}\up{0.83} & \textbf{83.37}\up{1.70} \\
\midrule
\midrule
\textbf{Ratio (2$\times$Augmentation $>$ Baseline)} & \multicolumn{2}{c}{7/37} & \multicolumn{2}{c}{28/37} & \multicolumn{2}{c}{16/37} & \multicolumn{2}{c}{24/37} \\
\textbf{Ratio (ASR-driven $>$ MT-driven)} & \multicolumn{2}{c}{19/36} & \multicolumn{2}{c}{20/36} & \multicolumn{2}{c}{23/36} & \multicolumn{2}{c}{22/36} \\
\bottomrule
\end{tabular}}
\vspace{-0.6cm}
\end{table*}

\vspace{-0.1cm}
\subsection{AD detection accuracy with TTS-based DA}
\vspace{-0.1cm}
\label{subsec:ad_detection_accuracy}
Table~\ref{tab:tts_aug_2x_results} compares the AD detection accuracy of the baseline (trained only on original training set) against models trained on $2\times$ augmented sets using various TTS models and text sources.  From this table, we observe that:
\textit{\textbf{1)}} CS-Cond models demonstrate superior augmentation efficacy compared to their pretrained counterparts.
Specifically, across all 37 text sources (MT + 36 ASR), CS-Cond CosyVoice2 surpasses the baseline in 28 cases (28/37), significantly outperforming the 7/37 of pretrained CosyVoice2. Similarly, CS-Cond F5-TTS achieves 24/37 compared to 16/37 for pretrained F5-TTS. This demonstrates that the speech generated by our CS-Cond TTS models is more effective for DA.
\textit{\textbf{2)}} ASR-driven augmentation frequently outperforms its MT-driven counterpart. For each TTS model, more than half of the 36 ASR-driven configurations yield higher accuracy than using MT (specifically 19/36, 20/36, 23/36, and 22/36). We attribute this to the nature of the text sources: MT provides perfect transcripts, resulting in synthetic speech that highly overlaps linguistically with the original speech. In contrast, ASR transcripts contain non-random recognition errors that likely reflect pathological acoustic features of AD patients (e.g., articulation slurring). These errors are preserved during speech synthesis, thereby increasing the diversity of the training data and enhancing AD detection performance.

\begin{table}[b!]
 \vspace{-0.7cm}
\centering
\caption{Effectiveness of TTA on AD detection accuracy (\%).}
\label{tab:test_time_aug}
\renewcommand{\arraystretch}{0.1} 
\setlength{\tabcolsep}{3pt} 
\resizebox{0.7\linewidth}{!}{
\begin{tabular}{@{}ccc@{}}
\toprule
\textbf{TTS text source}  & \textbf{w/o TTA} & \textbf{w/ TTA} \\ 
\midrule
Fine-tuned w2v960 & 84.17 & 85.42\up{1.25} \\
Fine-tuned w2v960 large lv & 85.00 & \textbf{85.83}\up{0.83} \\
Fine-tuned whisper large v3 & 84.58 & 85.42\up{0.84} \\
\midrule
Average & 84.58 & 85.56\up{0.98} \\ 
\bottomrule
\end{tabular}
}
\end{table}

\vspace{-0.1cm}
\subsection{Impact of augmentation factor on AD detection}
\vspace{-0.1cm}
We investigated the impact of various augmentation factors ($1\times$ to $4\times$) on AD detection accuracy by employing the Intra-Class Cross-Synthesis strategy. The results are illustrated in Figure~\ref{fig:aug_factor}. We observe that the performance follows an inverted-U curve, with the optimal range lying between $1.5\times$ and $2.5\times$. On average, the best performance is achieved at an augmentation factor of $2\times$ (a 1:1 mixture of original and synthetic data). Excessive augmentation degrades performance. This is likely because, while the synthetic speech is realistic, it inevitably contains artifacts. An excessively high proportion of synthetic data causes the detection model to overfit to the generative features of the TTS system rather than the genuine pathological characteristics.

\vspace{-0.1cm}
\subsection{Effectiveness of test-time augmentation (TTA)}
\vspace{-0.1cm}
Finally, we applied TTA to the three ASR text sources from Figure~\ref{fig:aug_factor} under the $2\times$ augmentation setting. The results are presented in Table~\ref{tab:test_time_aug}. We observe that after implementing TTA, the accuracy for all configurations improved by approximately 1\%, ultimately achieving a best accuracy of 85.83\%.
\vspace{-0.1cm}
\subsection{Comparison with traditional DA and previous studies}
\vspace{-0.1cm}
Table~\ref{tab:comparison_sota} presents a comprehensive comparison. On the left, we observe that traditional DA methods (e.g., noise addition, time stretching) yield only marginal improvements of 0.4\%--0.8\% or even a 2.5\% degradation (pitch shifting) over the baseline. On the right, compared to previous audio-only studies~\cite{gao2025leveraging,liu24f_interspeech,guo2023exploring}, our proposed CoSTA achieves a significantly higher accuracy of 85.83\%, surpassing the baseline by 4.16\% and demonstrating the effectiveness of CS-Cond TTS for DA.

\begin{table}[t!]
\centering
\caption{Comparison with traditional DA and previous studies.}
\label{tab:comparison_sota}
\renewcommand{\arraystretch}{1} 
\setlength{\tabcolsep}{4pt} 
\resizebox{1\linewidth}{!}{
\begin{tabular}{lc|lc}
\toprule
\textbf{Method (Traditional DA)} & \textbf{Accuracy (\%)} & \textbf{Method (Previous Studies)} & \textbf{Accuracy (\%)} \\
\midrule
Baseline (WavLM-based) & 81.67 & Whisper + MLP \cite{gao2025leveraging} & 79.17 \\
\quad + Noise Addition & 82.50\up{0.83} & Wav2Vec2 + Linear \cite{liu24f_interspeech} & 80.83 \\
\quad + Pitch Shifting & 79.17\down{2.5} &   AW-HuBERT \cite{guo2023exploring} & 81.67   \\
\quad + Time Stretching & 82.08\up{0.41} & \textbf{CoSTA (Ours)} & \textbf{85.83}\up{4.16}  \\
\bottomrule
\end{tabular}
}
\vspace{-0.8cm}
\end{table}

\vspace{-0.2cm}
\section{Conclusions}
\label{sec:Conclusions}
\vspace{-0.1cm}
In this paper, we proposed CoSTA, a novel DA framework that leverages CS-Cond TTS to address the inherent data scarcity challenge in speech-based AD detection. Our extensive experiments yield three key insights. First, explicitly conditioning TTS models on cognitive states enables the synthesis of speech with realistic pathological characteristics, significantly enhancing downstream augmentation utility compared to standard pretrained TTS. Second, ASR-driven augmentation frequently outperforms MT-driven augmentation. This suggests that non-random ASR errors likely capture diagnostically relevant linguistic perturbations that are effectively preserved during synthesis, thereby enriching the diversity and robustness of the augmented training set. Finally, by integrating TTA, CoSTA achieves an audio-only accuracy of 85.83\% on the ADReSS test set, surpassing the unaugmented baseline by 4.16\% and outperforming prior methods. Future work will explore cross-lingual pathological speech synthesis and leverage synthesized data to train unified understanding models for AD speech.

\section{Acknowledgments}
This work is supported in part by the National Social Science Foundation of China (Grant No. 23AYY012) and by the Supercomputing Center of the University of Science and Technology of China.

\section{Generative AI Use Disclosure}
Generative AI tools were used only for minor improvements in language and presentation. 
No AI system was used to generate, modify, or interpret the scientific content of this manuscript.

\bibliographystyle{IEEEtran}
\bibliography{mybib}

\end{document}